\documentclass{article}
\usepackage{spconf,amsmath,amsfonts,graphicx}
\usepackage{amssymb,textcomp,mathtools}
\usepackage{bm,upgreek,algorithm}
\usepackage{multirow, booktabs, hhline, array}
\usepackage{cite, url, makecell, setspace}
\usepackage{acronym}
\usepackage{subcaption}
\usepackage{xcolor}

\pdfoutput=1

\RequirePackage{doi}

\def\thline{\noalign{\hrule height 1.0pt}}

\renewcommand{\vec}[1]{\bm{\mathrm{#1}}}

\acrodef{ASR}[ASR]{automatic speech recognition}
\acrodef{FaSNet}[FaSNet]{filter-and-sum network}
\acrodef{TD}[TD]{time-domain}
\acrodef{MWF}[MWF]{multi-channel Wiener filter}
\acrodef{FaS}[FaS]{filter-and-sum}
\acrodef{MVDR}[MVDR]{minimum variance distortionless response}
\acrodef{MB}[MB]{mask-based}
\acrodef{GEV}[GEV]{generalized eigenvalue}
\acrodef{IBM}[IBM]{ideal binary mask}
\acrodef{FD}[FD]{frequency-domain}
\acrodef{T-F}[T-F]{time-frequency}
\acrodef{SDW-MWF}[SDW-MWF]{speech distortion weighted MWF} 
\acrodef{SNR}[SNR]{source-to-noise ratio}
\acrodef{DRR}[DRR]{direct-to-reverberant ratio}
\acrodef{NCC}[NCC]{normalized cross-correlation}
\acrodef{TCN}[TCN]{temporal convolutional network}
\acrodef{SI-SNR}[SI-SNR]{scale-invariant signal-to-noise ratio}
\acrodef{SI-MSE}[SI-MSE]{scale-invariant mean-square-error}
\acrodef{PIT}[PIT]{Permutation invariant training}
\acrodef{DOA}[DOA]{direction of arrival}
\acrodef{TDOA}[TDOA]{time difference of arrival}
\acrodef{ESE}[ESE]{Echoic noisy speech enhancement}
\acrodef{ESS}[ESS]{Echoic noisy speech separation }
\acrodef{RIR}[RIR]{room impulse response}
\acrodef{BN}[BN]{batch normalization}
\acrodef{OC}[OC]{open-condition}
\acrodef{CC}[CC]{close-condition}
\acrodef{WER}[WER]{word error rate}
\acrodef{RWERR}[RWERR]{relative word error rate reduction}

\title{FaSNet: LOW-LATENCY ADAPTIVE BEAMFORMING FOR MULTI-MICROPHONE AUDIO PROCESSING}

\twoauthors
  {Yi Luo*, \,Cong Han, \, Nima Mesgarani}
	{Columbia University\\
	Department of Electrical Engineering\\
	USA}
  {Enea Ceolini*\thanks{* These authors contributed equally to this work.}, \,Shih-Chii Liu}
	{University of Zurich and ETH Zurich\\
	Institute of Neuroinformatics\\
	Switzerland}

\begin{document}
\ninept
\maketitle

\section{Abstract}
\label{sec:abstract}

Beamforming has been extensively investigated for multi-channel audio processing tasks. Recently, learning-based beamforming methods, sometimes called \textit{neural beamformers}, have achieved significant improvements in both signal quality (e.g. signal-to-noise ratio (SNR)) and speech recognition (e.g. word error rate (WER)). Such systems are generally non-causal and require a large context for robust estimation of inter-channel features, which is impractical in  applications requiring low-latency responses. In this paper, we propose filter-and-sum network (FaSNet), a time-domain, filter-based beamforming approach suitable for low-latency scenarios. FaSNet has a two-stage system design that first learns frame-level time-domain adaptive beamforming filters for a selected reference channel, and then calculate the filters for all remaining channels. The filtered outputs at all channels are summed to generate the final output. Experiments show that despite its small model size, FaSNet is able to outperform several traditional oracle beamformers with respect to scale-invariant signal-to-noise ratio (SI-SNR) in reverberant speech enhancement and separation tasks. Moreover, when trained with a frequency-domain objective function on the CHiME-3 dataset, FaSNet achieves 14.3\% relative word error rate reduction (RWERR) compared with the baseline model. These results show the efficacy of FaSNet particularly in reverberant and noisy signal conditions. 

\begin{keywords}
Beamforming, multi-channel, audio processing, deep learning, low-latency
\end{keywords}

\section{Introduction}
\label{sec:intro}

\textit{Beamforming}, also known as spatial filtering, is a powerful microphone array processing technique that extracts the signal-of-interest in a particular direction and reduces the effect of noise and reverberation from a multi-channel signal \cite{gannot2017consolidated}. With the dominance of deep learning methods in almost all audio processing tasks, deep learning-based beamformers, sometimes called \textit{neural beamformers}, have also proven effective especially when jointly trained together with backend models for tasks such as \ac{ASR} \cite{xiao2016study}.

Most neural beamformers can be broadly categorized into three main categories. The first category, which we refer to as the \textit{filtering-based} (FB) approach, aims at learning a set of beamforming filters to perform \ac{FaS} beamforming in either the time domain \cite{sainath2015speaker, li2016neural, sainath2017multichannel} or frequency domain \cite{xiao2016deep, xiao2016beamforming, meng2017deep, jo2018estimation}. \ac{FaS} beamforming applies the beamforming filters to each channel and then sums them up to generate a single-channel output, within which the filters can be either fixed or adaptive depending on the model design. The second category, which we refer to as the \textit{masking-based} (MB) beamforming, estimates the \ac{FaS} beamforming filters in frequency domain by estimating \ac{T-F} masks for the sources of interest \cite{heymann2015blstm, heymann2016neural, erdogan2016multi, erdogan2016improved, xiao2017time, ochiai2017multichannel, ochiai2017unified, pfeifenberger2017dnn, boeddeker2017optimizing, heymann2017beamnet, zhang2017speech, boeddeker2018exploring, matsui2018online, heymann2018performance}. The \ac{T-F} masks specify the dominance of each \ac{T-F} bin and are used to calculate the spatial covariance features required to obtain optimal weights for beamformers such as \ac{MVDR} \cite{capon1969high} and \ac{GEV} beamformer \cite{warsitz2007blind}. The third category, which we refer to as the \textit{regression-based} (RB) approach, implicitly incorporates beamforming within a neural network without explicitly generating the beamforming filters \cite{stoller2018wave, grais2018raw}. In this framework, the input channels are directly passed to a neural network (typically a convolutional neural network) and the training objective is to learn a mapping between the multi-channel inputs and the target source of interest. The beamforming operation is thus assumed to be implicitly included in the mapping function defined by the model. Other methods that are not part of these three main categories include ones that combine neural networks and beamforming in different ways. For example, \cite{qian2018deep} uses a single-channel speech enhancement network to first estimate the source of interest and then applies time-domain Wiener-filtering based beamforming.

Previous studies have shown that frequency-domain neural beamformers significantly outnumber time-domain neural beamformers for several reasons. First, neural beamformers are typically designed and applied to \ac{ASR} tasks in which frequency-domain methods are still the most common approaches. Second, frequency-domain beamformers are known to be more robust and effective than time-domain beamformers in various tasks \cite{lockwood2004performance, hamid2014performance}. However, in applications and devices where online, low-latency processing is required, frequency-domain methods have the disadvantage that the frequency resolution and the input signal length needed for a reasonable performance might result in a large, perceivable system latency. For example, \ac{MB} beamforming methods rely on the efficacy of the mask estimation network whose performance will typically degrade in online or causal scenarios \cite{matsui2018online, boeddeker2018exploring}.

To address the limitation of previous neural beamformers, here we propose \ac{FaSNet}, a time-domain adaptive \ac{FaS} beamforming framework suitable for realtime, low-latency applications. \ac{FaSNet} consists of two stages where the first stage estimates the beamforming filter for a selected reference channel, and the second stage utilizes the output from the first stage to estimate beamforming filters for all remaining channels. The input for both stages consists of the target channel to be beamformed as well as the use of the \ac{NCC} between channels as the inter-channel feature. Both stages make use of the temporal convolutional networks (\ac{TCN} \cite{luo2018tasnet}) for low-resource, low-latency processing. Moreover, depending on the actual task to solve, the training objective of \ac{FaSNet} can be either a signal-level criterion (e.g. \ac{SNR}) or \ac{ASR}-level criterion (e.g. mel-spectrogram), which makes \ac{FaSNet} a flexible framework for various scenarios.

The remainder of the paper is organized as follows. We introduce the proposed \ac{FaSNet} model in Section~\ref{sec:FaSNet}, describe the experiment configurations in Section~\ref{sec:exp}, report the experiment results in Section~\ref{sec:result}, and conclude the paper in Section~\ref{sec:conclusion}.

\section{Filter-and-sum Network (FaSNet)}
\label{sec:FaSNet}
\begin{figure*}[!htp]
	\small
	\centering
	\includegraphics[width=1.5\columnwidth]{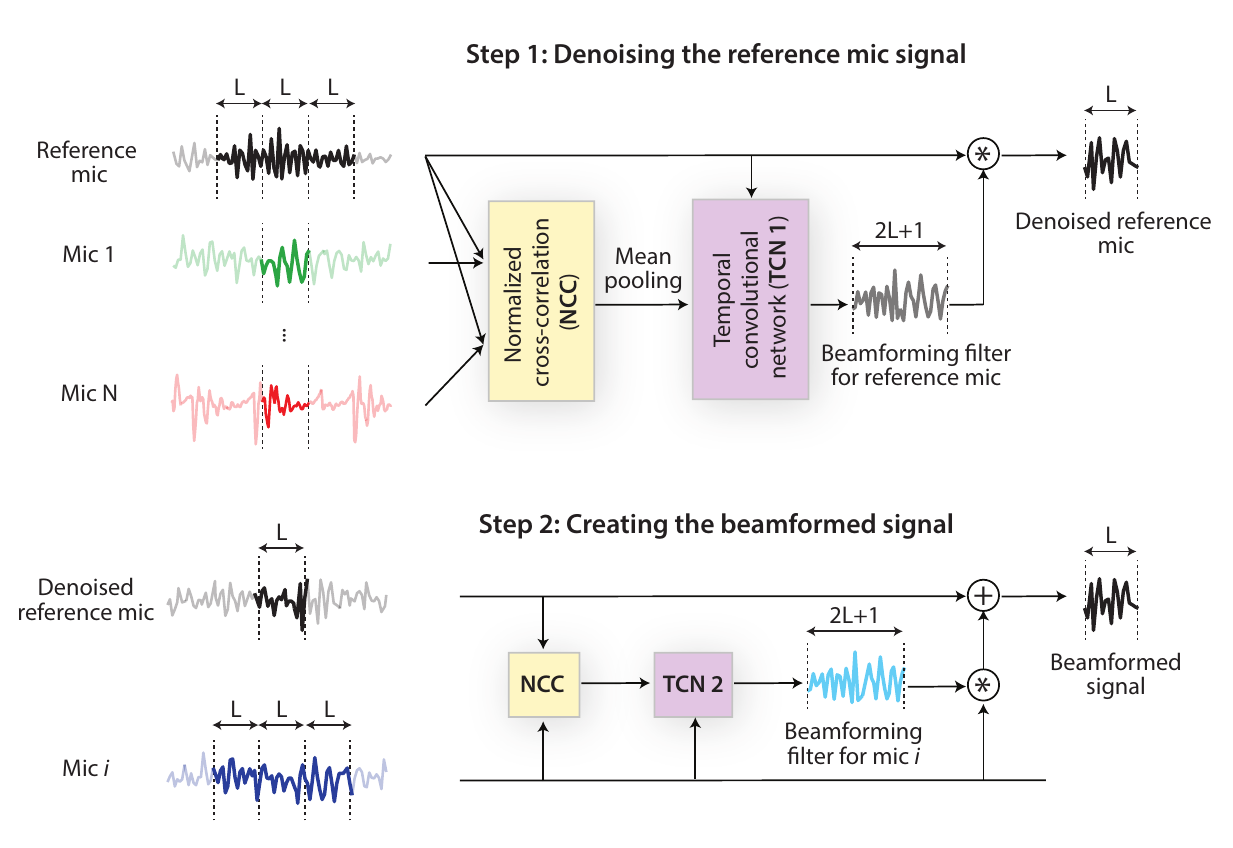}
	\caption{System flowchart for the proposed \ac{FaSNet} system. The first stage estimates the frame-level beamforming filters for the reference microphone based on the normalized correlation correlation coefficient (NCC) feature, and the second stage uses the cleaned reference microphone signal to estimate the beamforming filters for all remaining microphones. Cosine similarity is used as the NCC feature, and the temporal convolutional network (TCN) is selected as the filter estimation module.}
	\label{fig:flowchart}
\end{figure*}

\subsection{Problem definition}

The problem of time-domain \ac{FaS} beamforming is defined as estimating a set of time-domain filters for a microphone array of $N \geq 2$ microphones, such that the summation of the filtered signals of all microphones provides the best estimation of a signal of interest in a selected reference microphone. We first split the signals $\vec{x}^i$ at each microphone into frames of $L$ samples with a hop size of $H \in [0, L-1]$ samples
\begin{align}
	\vec{x}_t^i = \vec{x}^i[tH:tH+L-1], \quad t \in \mathbb{Z}, \quad i=1,\ldots, N
\end{align}
where $t$ is the frame index, $i$ is the index of the microphone and the operation $\vec{x}[a:b]$ selects the values of vector $\vec{x}$ from index $a$ to index $b$. To account for the \ac{TDOA} of the signal of interest at different microphones, the \ac{FaS} operation is applied on a context window around frame $t$ for each microphone to generate the beamformed output at frame $t$
\begin{align}
	\hat{\vec{y}}_t = \sum_{i=1}^N \vec{h}^i_t \circledast \hat{\vec{x}}^i_t
\end{align}
where $\hat{\vec{y}}_t \in \mathbb{R}^{1\times L}$ is the beamformed signal at frame $t$, $\hat{\vec{x}}^i_t = \vec{x}^i[tH -L:tH + 2L - 1] \in \mathbb{R}^{1\times 3L}$ is the context window around $\vec{x}_t$ for microphone $i$, $\vec{h}^i_t \in \mathbb{R}^{1\times (2L+1)}$ is the beamforming filter to be learned for microphone $i$, and $\circledast$ represents the convolution operation. For frames where $tH < L$ or $tH + 2L > l$ where $l$ is the total length of the signal, zero is padded to the context windows. The use of context window $\hat{\vec{x}}^i_t$ is to make sure the model can capture cross-microphone delays of $\pm L$ samples, since the directions of the sources are always unknown. As shown in \cite{li2016neural}, the use of the context window incorporates the estimation of cross-microphone delay into the learning process of $\vec{h}^i_t$. The problem of \ac{FaS} beamforming thus becomes estimating $\vec{h}^i_t$ given the observations of $\vec{x}^i$. For simplicity, we drop the frame index $t$ in the following discussions where there is no ambiguity.

\subsection{Reference channel processing}

The first stage in \ac{FaSNet} is to calculate the beamforming filter for the reference microphone which is randomly selected from the array (the first channel in all our experiments). Motivated by the GCC-PHAT feature \cite{knapp1976generalized, brandstein1997robust} in other frequency-domain beamformers and tasks such as \ac{DOA} and \ac{TDOA} estimation, we use frame-level normalized cross-correlation (NCC) as the inter-channel feature. To be specific, let $\hat{\vec{x}}^1 \in \mathbb{R}^{1\times 3L}$ be the context window of the signal in the reference microphone, and $\vec{x}^i \in \mathbb{R}^{1\times L}, i=2,\ldots,N$ be the corresponding center frame of all the other microphones with same index, then the \ac{NCC} feature, which we defined as the cosine similarity in our case, is calculated between $\hat{\vec{x}}^1$ and $\vec{x}^i$:
\begin{align}
	\begin{cases}
		\hat{\vec{x}}^1_j = \hat{\vec{x}}^1[j:j+L-1] \\
		f^i_j = \frac{\hat{\vec{x}}^1_j(\vec{x}^i)^T}{\left\|\hat{\vec{x}}^1_j\right\|_2 \left\|\vec{x}^i\right\|_2}
	\end{cases}, \quad j=1, \ldots, 2L+1
\end{align}
where $\vec{f}^i \in \mathbb{R}^{1\times (2L+1)}$ is the cosine similarity between reference microphone and microphone $i$. The \ac{NCC} feature contains both the \ac{TDOA} information and the content-dependent information of the signal of interest in the reference microphone and the other microphones. In order to combine the $N-1$ such features $\vec{f}^i, i=2,\ldots,N$ for all non-reference microphones in a permutation-free manner (i.e. independent from microphone indexes), we simply apply mean-pooling to them
\begin{align}
	\bar{\vec{f}}^i = \frac{1}{N-1}\sum_{i=2}^N \vec{f}^i
\end{align}

For channel-specific feature, a linear layer is applied on $\vec{x}^1 \in \mathbb{R}^{1\times L}$, the center frame of $\hat{\vec{x}}^1$, to create a $K$-dimensional embedding $\vec{R}^1 \in \mathbb{R}^{1\times K}$
\begin{align}
	\vec{R}^1 = \vec{x}^1 \vec{U}
\end{align}
where $\vec{U} \in \mathbb{R}^{L\times K}$ is the weight matrix. $\vec{R}^1$ is then concatenated with $\bar{\vec{f}}^i$ and passed to a \ac{TCN} with the same design as \cite{luo2018tasnet} followed by a gated output layer to generate $C$ beamforming filters $\vec{h}^1_c \in \mathbb{R}^{1\times (2L+1)}, \, c = 1, \ldots, C$ where $C$ is the number of sources of interest:
\begin{align}
	\vec{p}^1_{1,\ldots, C} &= \mathcal{H}_1\big([\vec{R}^1, \bar{\vec{f}}]\big) \\
	\vec{h}^1_c &= tanh(\vec{p}^1_c \vec{W}^1 + \vec{b}^1) \odot \sigma(\vec{p}^1_c \vec{V}^1 + \vec{q}^1)
\end{align}
where $\mathcal{H}_1(\cdot)$ is the mapping function defined by the TCN, $\vec{p}^1_c \in \mathbb{R}^{1\times K}$ is the output of TCN, $\vec{W}^1, \vec{V}^1 \in \mathbb{R}^{K\times (2L+1)}$ and $\vec{b}^1, \vec{q}^1 \in \mathbb{R}^{1\times (2L+1)}$ are weight and bias parameters of the gated output layer respectively, $tanh(\cdot)$ and $\sigma(\cdot)$ denote the hyperbolic tangent and sigmoid functions respectively, and $\odot$ represents the Hadamard product. $\vec{h}^1_c$ is then convolved with $\hat{\vec{x}}^1$ to generate the beamformed output of source $c$, $\hat{\vec{y}}^1_c \in \mathbb{R}^{1\times L}$, for the reference microphone
\begin{align}
	\hat{\vec{y}}^1_c = \hat{\vec{x}}^1 \circledast \vec{h}^1_c.
\end{align}

\subsection{Remaining channel processing}

The second stage in \ac{FaSNet} is to estimate the beamforming filters $\vec{h}^i_c, i=2,\ldots,N$ for all remaining microphones. Using the output of each estimated sources of interest from the first step $\hat{\vec{y}}^1_c$ as the cue, we apply the similar procedure as above to all the remaining microphones. For microphone $i$ with context window $\hat{\vec{x}}^i \in \mathbb{R}^{1\times 3L}$, the \ac{NCC} feature is calculated between it and $\hat{\vec{y}}^1_c$:
\begin{align}
	\begin{cases}
		\hat{\vec{x}}^i_j = \hat{\vec{x}}^i[j:j+L-1] \\
		g^i_{c, j} = \frac{\hat{\vec{x}}^i_j(\hat{\vec{y}}^1_c)^T}{\left\|\hat{\vec{x}}^i_j\right\|_2 \left\|\hat{\vec{y}}^1_c\right\|_2}
	\end{cases}, \quad j=1, \ldots, 2L+1
\end{align}

Another \ac{TCN} with its corresponding mapping function $\mathcal{H}_2(\cdot)$ is used to generate $\vec{h}^i$ given $\vec{g}^i_c \in \mathbb{R}^{1\times (2L+1)}$ and the linear transformation $\vec{R}^i = \vec{x}^i \vec{U}$:
\begin{align}
	\vec{p}^i_c &= \mathcal{H}_2\big([\vec{R}^i, \vec{g}^i_c]\big) \\
	\vec{h}^i_c &= tanh(\vec{p}^i_c \vec{W}^2 + \vec{b}^2) \odot \sigma(\vec{p}^i_c \vec{V}^2 + \vec{q}^2)
\end{align}
where $\vec{W}^2, \vec{V}^2 \in \mathbb{R}^{K\times (2L+1)}$ and $\vec{b}^2, \vec{q}^2 \in \mathbb{R}^{1\times (2L+1)}$ are weight and bias parameters of the gated output layer respectively. Note that all remaining microphones share the same \ac{TCN} and gated output layer. The filters $\vec{h}^i_c$ are then convolved with $\hat{\vec{x}}^i$ and summed up to $\hat{\vec{y}}^1_c$ to generate the final beamformed output of source $c$
\begin{align}
	\hat{\vec{y}_c} = \hat{\vec{y}}^1_c + \sum_{i=2}^N \hat{\vec{x}}^i \circledast \vec{h}^i_c
\end{align}
Finally, all segments in $\hat{\vec{y}}_c$ are transformed back to the full utterance $\vec{y}^*_c \in \mathbb{R}^{1\times l}$ through the overlap-and-add operation. Figure~\ref{fig:flowchart} shows the full diagram of the system.

\subsection{Combination with single-channel systems}

The output of \ac{FaSNet} can also be passed to any single-channel enhancement system for further performance improvement. As \ac{FaSNet} directly generates waveforms, the tandem system can still be trained end-to-end for either time-domain or frequency-domain objectives. Section~\ref{sec:result} shows that the tandem configuration leads to improved performance compared to both the single-channel system and \ac{FaSNet}-only system.

\subsection{Training objectives}

The training objective can either be in time- or frequency-domain depending on the actual task to solve. For tasks that take signal quality as an evaluation measure, we use the \ac{SI-SNR} \cite{luo2018speaker, Roux2019SDR} as the training objective:
\begin{align}
	\mathcal{L}_{obj} = \frac{1}{C}\sum_{c=1}^{C}\text{SI-SNR}(\vec{y}_c, \vec{y}^*_c).
\end{align}
For tasks where frequency-domain output is favored (e.g. \ac{ASR}), we use mel-spectrogram with \ac{SI-MSE} as the training objective:
\begin{align}
    &\begin{dcases}
    \vec{Y}_c = \left|\text{STFT}\left(\frac{\vec{y}_c}{\left\|\vec{y}_c\right\|_2}\right)\right| \\
    \vec{Y}^*_c = \left|\text{STFT}\left(\frac{\vec{y}^*_c}{\left\|\vec{y}^*_c\right\|_2}\right)\right| \\
    \end{dcases} \\
	&\mathcal{L}_{obj} = \frac{1}{C}\sum_{c=1}^{C}\text{MSE}(\vec{Y}_c\vec{M}, \vec{Y}^*_c\vec{M})
\end{align}
where $\vec{Y}_c, \vec{Y}^*_c \in \mathbb{R}^{T\times F}$ are the magnitude spectrograms of the target and estimated signals respectively, and $\vec{M} \in \mathbb{R}^{F\times D}$ is the mel-filterbank. Utterance-level permutation invariant training (uPIT) is always applied to address the output permutation problem \cite{Yu2017PIT, kolbaek2017multitalker}.


\section{Experiment configurations}
\label{sec:exp}
We evaluate the proposed \ac{FaSNet} on various types of multi-microphone audio processing tasks. To be specific, we design three different experiments: 
\begin{enumerate}
	\item \textbf{\ac{ESE}}: We jointly perform speech denoising and dereverberation in an echoic environment;
	\item \textbf{\ac{ESS}}: We separate the direct path of two speakers in a noisy, echoic environment;
    \item \textbf{Multichannel noisy \ac{ASR}}: We use the 3rd CHiME challenge dataset \cite{Barker2015TheT} for ASR task.
\end{enumerate}
The direct-path speech signals for all sources of interest are always used as the target.

\subsection{Data generation}

For \ac{ESE} and \ac{ESS} tasks, we assume a circular omni-directional microphone array with a maximum of 4 microphones evenly distributed. The diameter of the array is fixed to 10 cm. The positions of the sources (speakers and the noise) and the center of the microphone array are randomly sampled, with the constraint that all sources should be at least 0.5~m away from the room walls. The height for all sources is fixed to 1~m. We then simulate the \ac{RIR} filters with the image method \cite{allen1979image}, and specifically with the gpuRIR toolbox \cite{diaz2018gpurir}. We randomize the length and the width of the rooms within the range $\left[3, 8\right]$~m and fix the height to 3~m.

We then generate a dataset for simulated \ac{ESS} and \ac{ESE} scenarios with the TIMIT dataset \cite{garofolo1993darpa}. We first randomly split each speaker's utterance into 7 training, 2 validation and 1 test samples, and then generate the training, validation and test sets within the corresponding categories that contain 20000, 5000 and 3000 rooms respectively. Each room contains two speakers and one noise source, within which the noise is randomly sampled from first 80 samples in \cite{web100nonspeech} for training and validation sets and all 100 samples for the test set. For ESE, the relative \ac{SNR} between the speaker and the noise is randomly sampled between $\left[-5, 15 \right]$~dB. In ESS, the relative \ac{SNR} between the two speakers is randomly sampled between $\left[-5, 5 \right]$~dB and the noise is randomly sampled between $\left[-5, 15 \right]$~dB with respect to the low energy speaker. 

\subsection{Hyperparameters setting}

Table~\ref{tab:param} shows the symbols for the hyperparameters in the system. Each \ac{TCN} has an identical design to \cite{luo2018tasnet} and contains $R$ repeats of the 1-D convolutional blocks with $P$ blocks in each repeat. In all experiments, we set $R=2$ and $P=5$. The size of the 1-D convolutional kernel in each 1-D convolutional block is 3, and the input and hidden channels in each block are set to 64 and 320 respectively. The embedding dimension $K$ is set to 64. The number of parameters in each TCN is thus 0.76M. The effect of different frame length $L$ is discussed in Section~\ref{sec:exp-fasnet}. For tandem systems with a single-channel system for second-stage enhancement, we adopt the Conv-TasNet configuration \cite{luo2018tasnet} but change the masking layer into a direct regression layer. The model size of the single-output Conv-TasNet is 1.9M.
\begin{table}[!htbp]
	\small
	\centering
	\caption{Hyperparameters in the proposed system.}
	\vspace{0.2cm}
	\label{tab:param}
	\begin{tabular}{c|c}
		\thline
		Symbol & Description\\
		\thline
		$L$ & \thead{Frame size (in samples)}\\
		\hline
		$P$ & \thead{Number of convolutional blocks in each repeat in TCN}\\
		\hline
		$R$ & \thead{Number of repeats in TCN} \\
		\hline
		$K$ & \thead{Dimension of embeddings as well as the output of TCN}\\
		\thline
	\end{tabular}
\end{table}

In the \ac{ASR} and \ac{ESE} tasks, each \ac{TCN} estimates one beamforming filter at each frame, while in the \ac{ESS} task, each \ac{TCN} estimates two beamforming filters corresponding to the two speakers.

\subsection{Traditional beamformers}
\label{sec:cbf}
In order to show the advantages and performance of the proposed approach, \ac{FaSNet} is compared against a variety of classical beamformers. Both beamformers in the time- and frequency-domain are considered. The comparison on the time-domain beamformers  is carried out since it represents a fairer comparison to FaSNet which is also based on the time domain. This comparison is extended to more traditional and more robust frequency-domain beamformers which are vastly used in practice.

Four classes of beamformers are considered in the comparison. The first class is time-domain beamformers and comprises \ac{TD} \ac{MWF} and \ac{TD} \ac{MVDR} beamformers \cite{BaiTDBF2013}. The second class is \ac{FD} beamformers and considers the \ac{SDW-MWF} and \ac{MVDR} \cite{doclo2010acoustic} beamformers. For both these classes the eigendecomposition method is used in order to estimate the steering vector \cite{Sarradj2010} from the estimated spatial covariance. The third class comprises \ac{MB} beamformers, specifically \ac{MVDR} \cite{erdogan2016improved} and \ac{GEV} \cite{heymann2015blstm} beamformers. Both these \ac{MB} beamformers use the \ac{IBM} to estimate the beamforming filters. In the interest of space, the exact formulation of each of the beamformers is omitted. We direct the interested reader to the original formulations, which are referenced in table~\ref{tab:oracle_bf}, and the open source implementation\footnote{\url{https://pypi.org/project/beamformers/}}.

\section{Results and Discussion}
\label{sec:result}
\subsection{Benchmarking oracle beamforming techniques on signal quality measurement}

All the beamformers described in Section~\ref{sec:cbf} are tested on both \ac{ESE} and \ac{ESS} tasks and evaluated with signal quality measurement (i.e. SI-SNR). Both time- and frequency-domain beamformers use the full utterance to estimate the spatial covariance and consequently calculate the steering vector. Similarly, \ac{MB} beamformers use the oracle \ac{IBM} on the full utterance to calculate the spatial covariance matrices. Table~\ref{tab:oracle_bf} provides the SI-SNR improvement of all described traditional beamformers. Among the time-domain beamformers, TD-MVDR shows better performance in the \ac{ESS} task while TD-MWF is better in the \ac{ESE} task. Even though the differences are minimal, we confirm the statement in \cite{BaiTDBF2013} that for speech enhancement in time domain, \ac{MVDR} is typically better than \ac{MWF}. Among the frequency-domain beamformers, the SDW-MWF beamformer is significantly better than MVDR, given the fact that by design SDW-MWF also leads to better dereverberation. For mask-based beamformers, MVDR shows significantly better performance than GEV. This confirms the observation in \cite{heymann2015blstm} that GEV suffers from phase adjustment problems which can significantly decrease signal quality. The overall performance of frequency-domain beamformers is significantly better than time-domain beamformers especially with an increasing number of microphones \cite{hamid2014performance}.

As \ac{FaSNet} has a fixed receptive field defined by the TCNs, we also conduct another experiment where the spatial covariances and masks are estimated based on short segments of length $s \in \{100, 250, 500\}$~ms with two possible ways for the estimation: the spatial covariance is calculated over time for every non-overlapping segment, or only estimated once based on a segment randomly selected within the utterance. We found empirically that the two ways lead to results lacking significant differences, so here we only report the results from the former configuration. Table~\ref{tab:oracle_seg_size} shows the comparison of the best performing oracle beamformers in Table~\ref{tab:oracle_bf} with different segment sizes. For the widely-used MB-MVDR, a large enough receptive field is crucial for a reasonable performance which makes it harder to apply in rapid changing conditions.

\begin{table}[t]
	\small
	\centering
	\caption{Performance of oracle beamformers. SI-SNR improvement is reported. CC: close-condition (development) set. OC: open-condition (evaluation) set.}
	\label{tab:oracle_bf}
	\begin{tabular}{c|c|cc|cc}
		\thline
		\multirow{3}{*}{\thead{Method}} & \multirow{3}{*}{\thead{\# of \\ mics}} & \multicolumn{4}{c}{\thead{SI-SNRi (dB)}} \\
		\cline{3-6}
		 & & \multicolumn{2}{c}{ESE} & \multicolumn{2}{c}{ESS} \\
		 \cline{3-6}
		 & & CC & OC & CC & OC \\
		\thline
		\multirow{3}{*}{TD-MVDR~\cite{BaiTDBF2013}} & 2 & 2.1 & 2.6 & 3.2 & 3.4 \\
		 & 3 & 2.5 &2.9 & 4.2 & 4.3 \\
		 & 4 & \bf{2.8} & \bf{3.2} & 3.9 & 4.3 \\
		 \hline
		 \multirow{3}{*}{TD-MWF~\cite{BaiTDBF2013}} & 2 & 1.6  & 1.8& 3.1 & 3.2 \\
		 & 3 & 2.1 &2.5 & 3.9 &  4.2  \\
		 & 4 & 2.5 &2.7 & \bf{4.4} & \bf{4.5} \\
		 \thline
		\multirow{3}{*}{FD-MVDR~\cite{doclo2010acoustic}} & 2  & 2.1 & 2.0 & 2.1 &  2.1 \\
		 & 3 &3.2 &3.0 & 3.5 &  3.5  \\
		 & 4 &4.1 &3.9 & 4.6 &  4.5 \\
		 \hline
		 \multirow{3}{*}{FD-SDW-MWF~\cite{doclo2010acoustic}} & 2 & 3.7  & 3.6 & 3.3 &  3.1 \\
		 & 3 & 6.4 &6.2 & 5.9 &  5.9 \\
		 & 4 & \bf{8.1} & \bf{7.9} & \bf{7.6} & \bf{7.5} \\
		 \thline
		 \multirow{3}{*}{MB-MVDR~\cite{erdogan2016improved}} & 2 & 3.9 & 3.8 & 4.1 &  3.3 \\
		 & 3 & 5.8 &5.7 & 6.2 & 5.1 \\
		 & 4 & \bf{6.7} & \bf{6.6} & \bf{7.5} & \bf{6.3} \\
		 \hline
		 \multirow{3}{*}{MB-GEV~\cite{heymann2015blstm}} & 2 & -4.8 & -5.7 & -4.1 &  -3.6 \\
		 & 3 &0.8 & 0.9 & 1.1 & 0.6 \\
		 & 4 & 2.5 & 2.5& 2.9 &2.3 \\
		\thline
		
	\end{tabular}
\end{table}



\begin{table}[!htbp]
	\small
	\centering
	\caption{Performance of oracle beamformers with different segment sizes for spatial covariance estimation. SI-SNR improvement is reported only on the OC set.}
	\label{tab:oracle_seg_size}
	\begin{tabular}{c|c|c|cc}
		\thline
		\multirow{2}{*}{\thead{Method}} &
		\multirow{2}{*}{\thead{Segment \\ size (ms)}} &
		\multirow{2}{*}{\thead{\# of \\ mics}} & \multicolumn{2}{c}{\thead{SI-SNRi (dB)}} \\
		\cline{4-5}
		 & & & \multicolumn{1}{c}{ESE} & \multicolumn{1}{c}{ESS} \\
		 \cline{4-5}
		\thline
		 \multirow{3}{*}{FD-SDW-MWF \cite{doclo2010acoustic}} & \multirow{1}{*}{100} & 4 & 4.5 & 4.0 \\
		\cline{2-5}
		 & \multirow{1}{*}{250} & 4 & 5.7 & 5.3 \\
		 \cline{2-5}
		 & \multirow{1}{*}{500} & 4 & \bf{6.5} & \bf{6.1} \\
		 \thline
		 \multirow{3}{*}{MB-MVDR \cite{erdogan2016improved}} & \multirow{1}{*}{100} & 4 & -0.3 & -1.2 \\
		\cline{2-5}
		 & \multirow{1}{*}{250} & 4 & 3.0 & 2.7 \\
		 \cline{2-5}
		 & \multirow{1}{*}{500} & 4 & 4.7 & 4.6 \\
		\thline
	\end{tabular}
\end{table}


\begin{figure*}[!ht]
\small
\centering
   \begin{subfigure}[b]{\columnwidth}
   \includegraphics[width=\columnwidth]{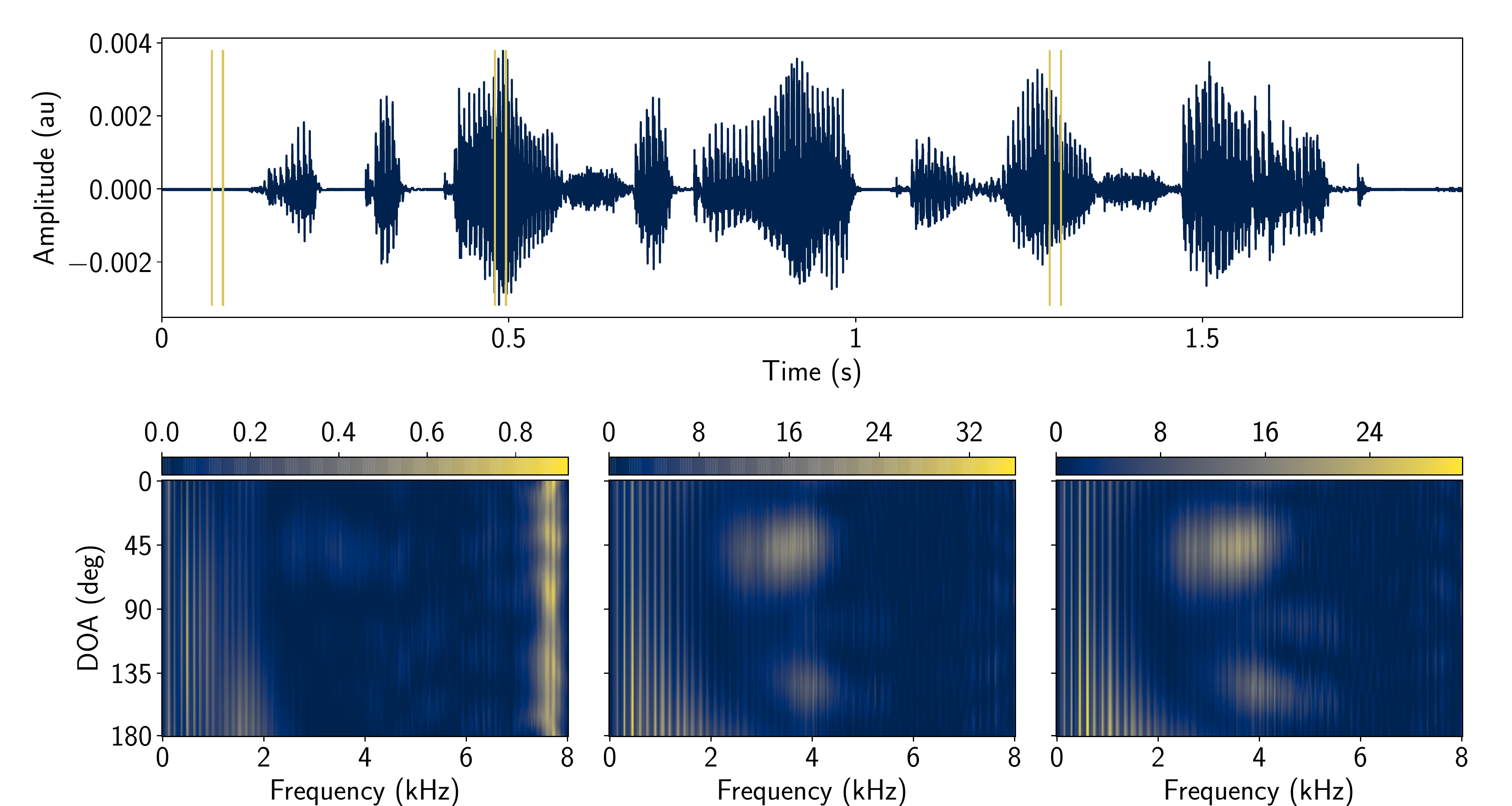}

\end{subfigure}
\begin{subfigure}[b]{\columnwidth}
   \includegraphics[width=\columnwidth]{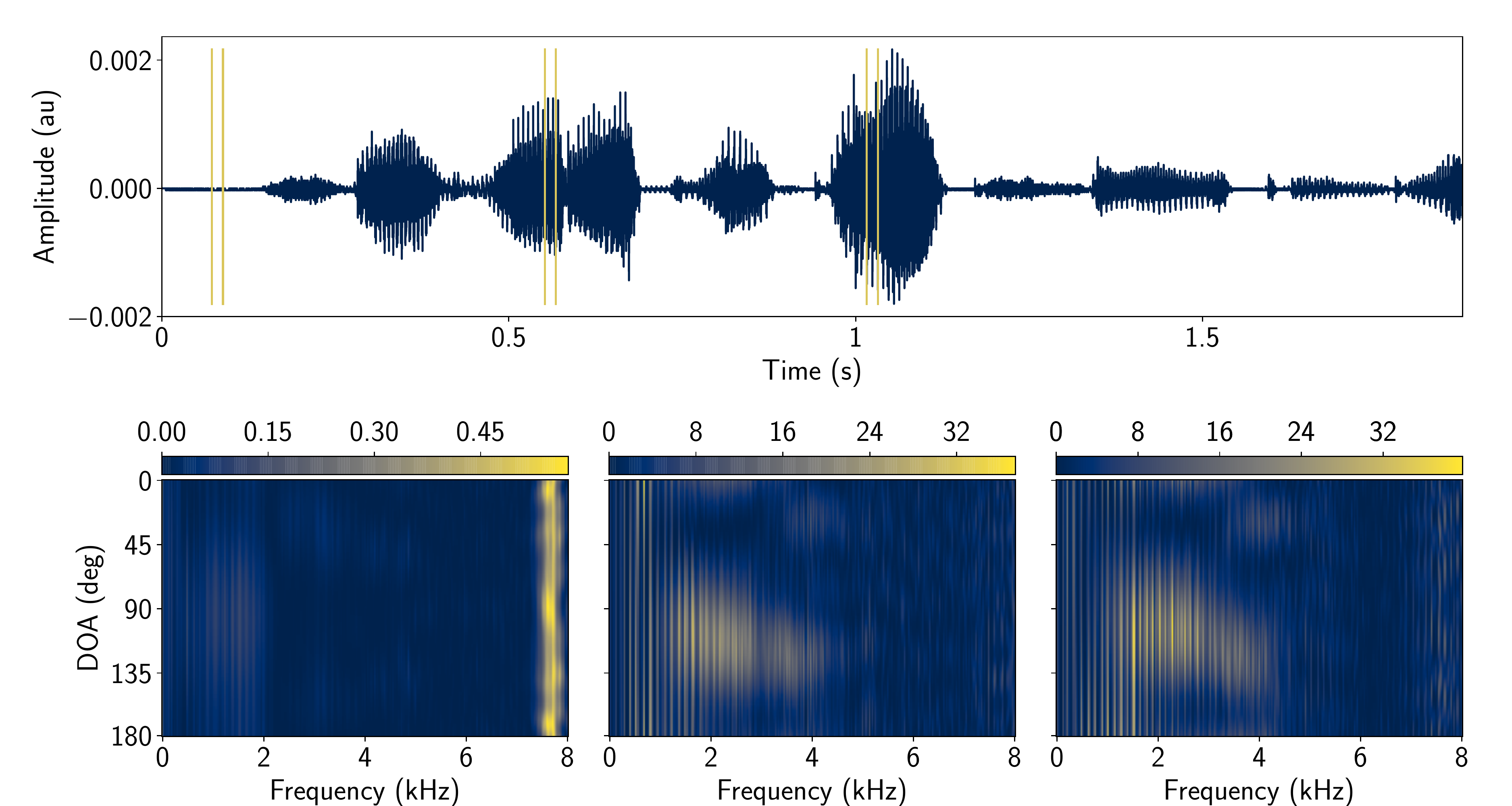}

\end{subfigure}
\caption{Beampattern examples for two different utterances in the ESS task.}
\label{fig:filter}
\end{figure*}

\subsection{Results of FaSNet on various tasks}
\label{sec:exp-fasnet}

 We first investigate the effect of frame size $L$ in \ac{FaSNet} while fixing other hyperparameters. Table~\ref{tab:fasnet_win} shows how different frame sizes affect the system performance. We can see that a longer frame size leads to constantly better performance, which is expected as higher frequency resolution can be achieved. As the system latency of \ac{FaSNet} is $2L$, tradeoff between performance and frame size needs to be considered for applications that strictly require low-latency processing. In this paper, we select the best performing system, i.e. $L=16$, for all following experiments. 
 

 \begin{table}[t]
	\small
	\centering
	\caption{Dependence of SI-SNR improvement on frame size for a 2-ch FaSNet in ESE task.
	}
	\vspace{0.2cm}
	\label{tab:fasnet_win}
	\begin{tabular}{c|cccc}
		\thline
		\multirow{2}{*}{} & \multicolumn{4}{c}{\thead{Frame (ms)}} \\
		\cline{2-5}
		 & 2 & 4 & 8 & 16 \\
		 \thline
		CC & 1.6 & 2.4 & 3.3 & \bf{4.0} \\
		\hline
		OC & 1.4 & 2.2 & 3.0 & \bf{3.7} \\
		\thline
	\end{tabular}
\end{table}

We then compare \ac{FaSNet} with a single-channel time-domain method, the Conv-TasNet \cite{luo2018tasnet}, on both tasks with various configurations. We choose this comparison as both models are based on the same TCN modules and have similar model design paradigm. Table~\ref{tab:fasnet_all} provides the comparison across different number of microphones and causality settings. We can observe that in a non-causal setting, \ac{FaSNet} achieves on par performance with the single-channel Conv-TasNet baseline of 4 microphones, while in a causal setting, it outperforms Conv-TasNet even with only 2 microphones. Moreover, adding a post single-channel enhancement network constantly improves the performance across almost all configurations on both tasks. The 4-channel tandem system is able to achieve on par performance with an MB-MVDR system with oracle IBM, and is significantly better than the segment-level oracle MB-MVDR. This shows that when comparing with frequency-domain beamformers which highly rely on a long segment for robust spatial covariance estimation, \ac{FaSNet} has better potential for low-latency processing on much shorter segments.

\begin{table}[!h]
	\small
	\centering
	\caption{Performance of FaSNet and tandem system in both ESE and ESS tasks.}
	\vspace{0.2cm}
	\label{tab:fasnet_all}
	\begin{tabular}{c|c|c|c|cc|cc}
		\thline
		\multirow{3}{*}{\thead{Method}} & \multirow{3}{*}{\thead{Model \\ size}} & \multirow{3}{*}{\thead{Causal}} & \multirow{3}{*}{\thead{\# of \\ mics}} & \multicolumn{4}{c}{\thead{SI-SNRi (dB)}} \\
		\cline{5-8}
		 & & & & \multicolumn{2}{c}{ESE} & \multicolumn{2}{c}{ESS} \\
		 \cline{5-8}
		 & & & & CC & OC & CC & OC \\
		\thline
		\multirow{2}{*}{Conv-TasNet} & \multirow{2}{*}{1.9M} & \texttimes & \multirow{2}{*}{1} & 5.3 & 5.0 & 4.1 & 4.1 \\
		\cline{3-3}\cline{5-8}
		& & \checkmark & & 3.5 & 3.3 & 2.7 & 2.6 \\
		\thline
		\multirow{6}{*}{FaSNet} & \multirow{6}{*}{1.5M} & \multirow{3}{*}{\texttimes} & 2 & 4.0 & 3.7 & 3.7 & 3.6 \\
		& & & 3 & 4.4 & 4.1 & 4.0 & 3.9 \\
		& & & 4 & \bf{5.3} & \bf{5.0} & \bf{4.7} & \bf{4.6} \\
		\cline{3-8}
		& & \multirow{3}{*}{\checkmark} & 2 & 3.8 & 3.5 & 3.2 & 3.1 \\
		& & & 3 & 4.1 & 3.8 & 3.5 & 3.4 \\
		& & & 4 & \bf{4.5} & \bf{4.3} & \bf{3.9} & \bf{3.8} \\ 
		\thline
		\multirow{6}{*}{Tandem} & \multirow{6}{*}{3.4M} & \multirow{3}{*}{\texttimes} & 2 & 5.8 & 5.5 & 4.5 & 4.4 \\
		& & & 3 & 5.4 & 5.0 & 5.5 & 5.5 \\
		& & & 4 & \bf{6.7} & \bf{6.4} & \bf{6.2} & \bf{6.1} \\
		\cline{3-8}
		& & \multirow{3}{*}{\checkmark} & 2 & 4.8 & 4.5 
		& 3.9 & 3.8\\
		& & & 3 & \bf{5.3} & \bf{5.0} & 4.1 & 4.0\\
		& & & 4 & 4.7 & 4.4 & \bf{4.5} & \bf{4.4}\\
		\thline
	\end{tabular}
\end{table}

We also evaluate \ac{FaSNet} on CHiME-3 dataset to investigate its potential as the front-end for speech recognition systems. Table~\ref{tab:fasnet_chime_sdr} shows the performance of \ac{FaSNet} with respect to signal quality measure. Two different training targets, the reverberant clean signal or the original clean signal, are applied during training. Note that the original clean source has an unknown shift with the oracle direct path signal in the reference microphone, so we adopt \textit{shift invariant training (SIT)} where we calculate the maximum SI-SNR between the system output and the original clean signal with $\pm 2$ ms of shift. We can see that \ac{FaSNet} is significantly better than the Conv-TasNet baseline with both targets, further proving its effectiveness on real-world recordings. 

Table~\ref{tab:fasnet_chime_wer} compares the \ac{WER} of \ac{FaSNet} and the official CHiME-3 baseline system on the recognition task. We use the officially provided DNN baseline recognizer as our backend ASR system, although more advanced systems with fully end-to-end training may further boost the performance. We can see from the table that when training with the original clean source as target and SI-SNR as objective, \ac{FaSNet} is able to achieve 9.3\% \ac{RWERR} compared with the MVDR baseline, and when training with the mel-spectrogram of the original clean signal as target with SI-MSE as objective, FaSNet achieves a 14.3\% \ac{RWERR}. This result proves that when training with a frequency-domain objective that favors ASR backends, \ac{FaSNet} can also serve as an effective ASR front-end.





\begin{table}[!htbp]
	\small
	\centering
	\caption{Performance of FaSNet on CHiME-3 evaluation dataset. SI-SNR improvement is reported.}
	\vspace{0.2cm}
	\label{tab:fasnet_chime_sdr}
	\begin{tabular}{c|c|c|c}
		\thline
        Target & Method & Causal & SI-SNRi (dB) \\ 
		\thline
		\multirow{3}{*}{Reverberant clean} & Conv-TasNet &  \texttimes & 8.7 \\
		\cline{2-4}
		& \multirow{2}{*}{FaSNet} & \texttimes & \bf{12.2}\\
 		\cline{3-4}
		& & \checkmark  & 10.6 \\
		\thline
		\multirow{3}{*}{Clean source} & Conv-TasNet &  \texttimes & 7.5 \\
		\cline{2-4}
		& \multirow{2}{*}{FaSNet} & \texttimes & \bf{11.6}\\
 		\cline{3-4}
		& & \checkmark  & 11.1 \\
		\thline
	\end{tabular}
\end{table}

\begin{table}[!htbp]
	\small
	\centering
	\caption{Performance of FaSNet on CHiME-3 evaluation dataset of real recordings. WER is reported.}
	\vspace{0.2cm}
	\label{tab:fasnet_chime_wer}
	\begin{tabular}{c|c|c}
		\thline
        Method & Target & WER (\%) \\ 
		\thline
		Noisy & -  & 32.53 \\
		\thline
		Baseline &  - & 32.48 \\
        \thline
		\multirow{3}{*}{FaSNet} & Reverberant clean & 32.23 \\
		& Clean source & 29.47 \\
		& Mel-spectrogram & \bf{27.89} \\
		\thline
	\end{tabular}
\end{table}

\subsection{Visualization of FaSNet filters}

To better understand the beampatterns of the time-domain filters generated by \ac{FaSNet}, Fig.~\ref{fig:filter} visualizes them for two example utterances in the ESS task. The figure shows the beampatterns estimated by FasNet at different frames of the utterances. The beampatterns are shown as a function of frequency and \ac{DOA}. As we can see, FasNet learns specific beampatterns which are content-dependent within each utterance, where different regions have different beampatterns. Specifically, nonspeech regions receive filters with null pattern for both utterances, further proving the adaptation ability of FaSNet across the utterance.
\vspace{-0.2cm}
\section{Conclusion}
\label{sec:conclusion}
In this paper, we proposed FaSNet, a time-domain adaptive beamforming method especially suitable for online, low-latency applications. FaSNet was designed as a two-stage system, where the first stage estimated the beamforming filter for a randomly selected reference microphone, and the second stage used the output of the first stage to calculate the filters for all the remaining microphones. FaSNet can also be concatenated with any other single-channel system for further performance improvement. Experimental results showed that FaSNet achieved better or on par performance than several oracle traditional beamformers on both echoic noisy speech enhancement (ESE) and echoic noisy speech separation (ESS) tasks. Moreover, when training with a frequency-domain objective that is favored by backend systems for speech recognition, FaSNet improved the word error rate on CHiME-3 by 14.3\% compared with a baseline model. Visualization on the beampatterns generated by FaSNet showed that it can estimate content-dependent adaptive filters for speech and nonspeech regions. Future work include investigations into the system performance on more diverse environments, e.g. with nonstationary speakers, and the evaluation of the system performance when trained with an ASR backend in a fully end-to-end manner.
\vspace{-0.2cm}
\section{Acknowledgments}
This work was partially supported by the EU H2020 grant No. 644732; the SNSF grant No. 20002$\_$1172553, a grant from the National Institute of Health, NIDCD, DC014279; a National Science Foundation CAREER Award; and the Pew Charitable Trusts.

\bibliographystyle{IEEEbib}
\bibliography{refs}

\begin{thebibliography}{10}

\bibitem{gannot2017consolidated}
Sharon Gannot, Emmanuel Vincent, Shmulik Markovich-Golan, and Alexey Ozerov,
\newblock ``A consolidated perspective on multimicrophone speech enhancement
  and source separation,''
\newblock {\em IEEE/ACM Transactions on Audio, Speech, and Language
  Processing}, vol. 25, no. 4, pp. 692--730, 2017.

\bibitem{xiao2016study}
Xiong Xiao, Chenglin Xu, Zhaofeng Zhang, Shengkui Zhao, Sining Sun, Shinji
  Watanabe, Longbiao Wang, Lei Xie, Douglas~L Jones, Eng~Siong Chng, et~al.,
\newblock ``A study of learning based beamforming methods for speech
  recognition,''
\newblock in {\em CHiME 2016 workshop}, 2016, pp. 26--31.

\bibitem{sainath2015speaker}
Tara~N Sainath, Ron~J Weiss, Kevin~W Wilson, Arun Narayanan, Michiel Bacchiani,
  et~al.,
\newblock ``Speaker location and microphone spacing invariant acoustic modeling
  from raw multichannel waveforms,''
\newblock in {\em Automatic Speech Recognition and Understanding (ASRU), 2015
  IEEE Workshop on}. IEEE, 2015, pp. 30--36.

\bibitem{li2016neural}
Bo~Li, Tara~N Sainath, Ron~J Weiss, Kevin~W Wilson, and Michiel Bacchiani,
\newblock ``Neural network adaptive beamforming for robust multichannel speech
  recognition.,''
\newblock in {\em Proc. Interspeech}, 2016, pp. 1976–--1980.

\bibitem{sainath2017multichannel}
Tara~N Sainath, Ron~J Weiss, Kevin~W Wilson, Bo~Li, Arun Narayanan, Ehsan
  Variani, Michiel Bacchiani, Izhak Shafran, Andrew Senior, Kean Chin, et~al.,
\newblock ``Multichannel signal processing with deep neural networks for
  automatic speech recognition,''
\newblock {\em IEEE/ACM Transactions on Audio, Speech, and Language
  Processing}, vol. 25, no. 5, pp. 965--979, 2017.

\bibitem{xiao2016deep}
Xiong Xiao, Shinji Watanabe, Hakan Erdogan, Liang Lu, John Hershey, Michael~L
  Seltzer, Guoguo Chen, Yu~Zhang, Michael Mandel, and Dong Yu,
\newblock ``Deep beamforming networks for multi-channel speech recognition,''
\newblock in {\em Acoustics, Speech and Signal Processing (ICASSP), 2016 IEEE
  International Conference on}. IEEE, 2016, pp. 5745--5749.

\bibitem{xiao2016beamforming}
Xiong Xiao, Shinji Watanabe, Eng~Siong Chng, and Haizhou Li,
\newblock ``Beamforming networks using spatial covariance features for
  far-field speech recognition,''
\newblock in {\em Signal and Information Processing Association Annual Summit
  and Conference (APSIPA), 2016 Asia-Pacific}. IEEE, 2016, pp. 1--6.

\bibitem{meng2017deep}
Zhong Meng, Shinji Watanabe, John~R Hershey, and Hakan Erdogan,
\newblock ``Deep long short-term memory adaptive beamforming networks for
  multichannel robust speech recognition,''
\newblock {\em arXiv preprint arXiv:1711.08016}, 2017.

\bibitem{jo2018estimation}
Moon~Ju Jo, Geon~Woo Lee, Jung~Min Moon, Choongsang Cho, and Hong~Kook Kim,
\newblock ``Estimation of mvdr beamforming weights based on deep neural
  network,''
\newblock in {\em Audio Engineering Society Convention 145}. Audio Engineering
  Society, 2018.

\bibitem{heymann2015blstm}
Jahn Heymann, Lukas Drude, Aleksej Chinaev, and Reinhold Haeb-Umbach,
\newblock ``Blstm supported gev beamformer front-end for the 3rd chime
  challenge,''
\newblock in {\em Automatic Speech Recognition and Understanding (ASRU), 2015
  IEEE Workshop on}. IEEE, 2015, pp. 444--451.

\bibitem{heymann2016neural}
Jahn Heymann, Lukas Drude, and Reinhold Haeb-Umbach,
\newblock ``Neural network based spectral mask estimation for acoustic
  beamforming,''
\newblock in {\em Acoustics, Speech and Signal Processing (ICASSP), 2016 IEEE
  International Conference on}. IEEE, 2016, pp. 196--200.

\bibitem{erdogan2016multi}
Hakan Erdogan, Tomoki Hayashi, John~R Hershey, Takaaki Hori, Chiori Hori,
  Wei-Ning Hsu, Suyoun Kim, Jonathan Le~Roux, Zhong Meng, and Shinji Watanabe,
\newblock ``Multi-channel speech recognition: Lstms all the way through,''
\newblock in {\em CHiME-4 workshop}, 2016.

\bibitem{erdogan2016improved}
Hakan Erdogan, John~R Hershey, Shinji Watanabe, Michael~I Mandel, and Jonathan
  Le~Roux,
\newblock ``Improved mvdr beamforming using single-channel mask prediction
  networks.,''
\newblock in {\em Proc. Interspeech}, 2016, pp. 1981--1985.

\bibitem{xiao2017time}
Xiong Xiao, Shengkui Zhao, Douglas~L Jones, Eng~Siong Chng, and Haizhou Li,
\newblock ``On time-frequency mask estimation for mvdr beamforming with
  application in robust speech recognition,''
\newblock in {\em Acoustics, Speech and Signal Processing (ICASSP), 2017 IEEE
  International Conference on}. IEEE, 2017, pp. 3246--3250.

\bibitem{ochiai2017multichannel}
Tsubasa Ochiai, Shinji Watanabe, Takaaki Hori, and John~R Hershey,
\newblock ``Multichannel end-to-end speech recognition,''
\newblock {\em arXiv preprint arXiv:1703.04783}, 2017.

\bibitem{ochiai2017unified}
Tsubasa Ochiai, Shinji Watanabe, Takaaki Hori, John~R Hershey, and Xiong Xiao,
\newblock ``Unified architecture for multichannel end-to-end speech recognition
  with neural beamforming,''
\newblock {\em IEEE Journal of Selected Topics in Signal Processing}, vol. 11,
  no. 8, pp. 1274--1288, 2017.

\bibitem{pfeifenberger2017dnn}
Lukas Pfeifenberger, Matthias Z{\"o}hrer, and Franz Pernkopf,
\newblock ``Dnn-based speech mask estimation for eigenvector beamforming,''
\newblock in {\em Acoustics, Speech and Signal Processing (ICASSP), 2017 IEEE
  International Conference on}. IEEE, 2017, pp. 66--70.

\bibitem{boeddeker2017optimizing}
Christoph Boeddeker, Patrick Hanebrink, Lukas Drude, Jahn Heymann, and Reinhold
  Haeb-Umbach,
\newblock ``Optimizing neural-network supported acoustic beamforming by
  algorithmic differentiation,''
\newblock in {\em Acoustics, Speech and Signal Processing (ICASSP), 2017 IEEE
  International Conference on}. IEEE, 2017, pp. 171--175.

\bibitem{heymann2017beamnet}
Jahn Heymann, Lukas Drude, Christoph Boeddeker, Patrick Hanebrink, and Reinhold
  Haeb-Umbach,
\newblock ``Beamnet: End-to-end training of a beamformer-supported
  multi-channel asr system,''
\newblock in {\em Acoustics, Speech and Signal Processing (ICASSP), 2017 IEEE
  International Conference on}. IEEE, 2017, pp. 5325--5329.

\bibitem{zhang2017speech}
Xueliang Zhang, Zhong-Qiu Wang, and DeLiang Wang,
\newblock ``A speech enhancement algorithm by iterating single-and
  multi-microphone processing and its application to robust asr,''
\newblock in {\em Acoustics, Speech and Signal Processing (ICASSP), 2017 IEEE
  International Conference on}. IEEE, 2017, pp. 276--280.

\bibitem{boeddeker2018exploring}
Christoph Boeddeker, Hakan Erdogan, Takuya Yoshioka, and Reinhold Haeb-Umbach,
\newblock ``Exploring practical aspects of neural mask-based beamforming for
  far-field speech recognition,''
\newblock in {\em 2018 IEEE International Conference on Acoustics, Speech and
  Signal Processing (ICASSP)}. IEEE, 2018, pp. 6697--6701.

\bibitem{matsui2018online}
Yutaro Matsui, Tomohiro Nakatani, Marc Delcroix, Keisuke Kinoshita, Nobutaka
  Ito, Shoko Araki, and Shoji Makino,
\newblock ``Online integration of dnn-based and spatial clustering-based mask
  estimation for robust mvdr beamforming,''
\newblock in {\em 2018 16th International Workshop on Acoustic Signal
  Enhancement (IWAENC)}. IEEE, 2018, pp. 71--75.

\bibitem{heymann2018performance}
Jahn Heymann, Michiel Bacchiani, and Tara~N Sainath,
\newblock ``Performance of mask based statistical beamforming in a smart home
  scenario,''
\newblock in {\em 2018 IEEE International Conference on Acoustics, Speech and
  Signal Processing (ICASSP)}. IEEE, 2018, pp. 6722--6726.

\bibitem{capon1969high}
Jack Capon,
\newblock ``High-resolution frequency-wavenumber spectrum analysis,''
\newblock {\em Proceedings of the IEEE}, vol. 57, no. 8, pp. 1408--1418, 1969.

\bibitem{warsitz2007blind}
Ernst Warsitz and Reinhold Haeb-Umbach,
\newblock ``Blind acoustic beamforming based on generalized eigenvalue
  decomposition,''
\newblock {\em IEEE Transactions on audio, speech, and language processing},
  vol. 15, no. 5, pp. 1529--1539, 2007.

\bibitem{stoller2018wave}
Daniel Stoller, Sebastian Ewert, and Simon Dixon,
\newblock ``Wave-u-net: A multi-scale neural network for end-to-end audio
  source separation,''
\newblock {\em arXiv preprint arXiv:1806.03185}, 2018.

\bibitem{grais2018raw}
Emad~M Grais, Dominic Ward, and Mark~D Plumbley,
\newblock ``Raw multi-channel audio source separation using multi-resolution
  convolutional auto-encoders,''
\newblock {\em arXiv preprint arXiv:1803.00702}, 2018.

\bibitem{qian2018deep}
Kaizhi Qian, Yang Zhang, Shiyu Chang, Xuesong Yang, Dinei Florencio, and Mark
  Hasegawa-Johnson,
\newblock ``Deep learning based speech beamforming,''
\newblock {\em arXiv preprint arXiv:1802.05383}, 2018.

\bibitem{lockwood2004performance}
Michael~E Lockwood, Douglas~L Jones, Robert~C Bilger, Charissa~R Lansing,
  William~D O’Brien~Jr, Bruce~C Wheeler, and Albert~S Feng,
\newblock ``Performance of time-and frequency-domain binaural beamformers based
  on recorded signals from real rooms,''
\newblock {\em The Journal of the Acoustical Society of America}, vol. 115, no.
  1, pp. 379--391, 2004.

\bibitem{hamid2014performance}
Umar Hamid, Rahim~Ali Qamar, and Kashif Waqas,
\newblock ``Performance comparison of time-domain and frequency-domain
  beamforming techniques for sensor array processing,''
\newblock in {\em Proceedings of 2014 11th International Bhurban Conference on
  Applied Sciences \& Technology (IBCAST) Islamabad, Pakistan, 14th-18th
  January, 2014}. IEEE, 2014, pp. 379--385.

\bibitem{luo2018tasnet}
Yi~Luo and Nima Mesgarani,
\newblock ``Tasnet: Surpassing ideal time-frequency masking for speech
  separation,''
\newblock {\em arXiv preprint arXiv:1809.07454}, 2018.

\bibitem{knapp1976generalized}
Charles Knapp and Glifford Carter,
\newblock ``The generalized correlation method for estimation of time delay,''
\newblock {\em IEEE transactions on acoustics, speech, and signal processing},
  vol. 24, no. 4, pp. 320--327, 1976.

\bibitem{brandstein1997robust}
Michael~S Brandstein and Harvey~F Silverman,
\newblock ``A robust method for speech signal time-delay estimation in
  reverberant rooms,''
\newblock in {\em Acoustics, Speech, and Signal Processing, 1997. ICASSP-97.,
  1997 IEEE International Conference on}. IEEE, 1997, vol.~1, pp. 375--378.

\bibitem{luo2018speaker}
Yi~Luo, Zhuo Chen, and Nima Mesgarani,
\newblock ``Speaker-independent speech separation with deep attractor
  network,''
\newblock {\em IEEE/ACM Transactions on Audio, Speech, and Language
  Processing}, vol. 26, no. 4, pp. 787--796, 2018.

\bibitem{Roux2019SDR}
Jonathan Le~Roux, Scott Wisdom, Hakan Erdogan, and John~R. Hershey,
\newblock ``Sdr – half-baked or well done?,''
\newblock in {\em 2019 IEEE International Conference on Acoustics, Speech and
  Signal Processing (ICASSP)}, May 2019, pp. 626--630.

\bibitem{Yu2017PIT}
D.~{Yu}, M.~{Kolbæk}, Z.~{Tan}, and J.~{Jensen},
\newblock ``Permutation invariant training of deep models for
  speaker-independent multi-talker speech separation,''
\newblock in {\em 2017 IEEE International Conference on Acoustics, Speech and
  Signal Processing (ICASSP)}, March 2017, pp. 241--245.

\bibitem{kolbaek2017multitalker}
Morten Kolb{\ae}k, Dong Yu, Zheng-Hua Tan, Jesper Jensen, Morten Kolbaek, Dong
  Yu, Zheng-Hua Tan, and Jesper Jensen,
\newblock ``Multitalker speech separation with utterance-level permutation
  invariant training of deep recurrent neural networks,''
\newblock {\em IEEE/ACM Transactions on Audio, Speech and Language Processing
  (TASLP)}, vol. 25, no. 10, pp. 1901--1913, 2017.

\bibitem{Barker2015TheT}
Jon Barker, Ricard Marxer, Emmanuel Vincent, and Shinji Watanabe,
\newblock ``The third ‘chime’ speech separation and recognition challenge:
  Dataset, task and baselines,''
\newblock {\em 2015 IEEE Workshop on Automatic Speech Recognition and
  Understanding (ASRU)}, pp. 504--511, 2015.

\bibitem{allen1979image}
Jont~B Allen and David~A Berkley,
\newblock ``Image method for efficiently simulating small-room acoustics,''
\newblock {\em The Journal of the Acoustical Society of America}, vol. 65, no.
  4, pp. 943--950, 1979.

\bibitem{diaz2018gpurir}
David Diaz-Guerra, Antonio Miguel, and Jose~R Beltran,
\newblock ``gpurir: A python library for room impulse response simulation with
  gpu acceleration,''
\newblock {\em arXiv preprint arXiv:1810.11359}, 2018.

\bibitem{garofolo1993darpa}
John~S Garofolo, Lori~F Lamel, William~M Fisher, Jonathan~G Fiscus, and David~S
  Pallett,
\newblock ``Timit acoustic-phonetic continous speech corpus cd-rom. nist speech
  disc 1-1.1,''
\newblock {\em NASA STI/Recon technical report n}, vol. 93, 1993.

\bibitem{web100nonspeech}
Guoning Hu,
\newblock ``100 {N}onspeech {S}ounds,'' {\small
  \url{http://web.cse.ohio-state.edu/pnl/corpus/HuNonspeech/HuCorpus.html}}.

\bibitem{BaiTDBF2013}
Mingsian Bai, Jeong-Guon Ih, and Jacob Benesty,
\newblock {\em Time-Domain MVDR Array Filter for Speech Enhancement},
  chapter~7, pp. 287--314,
\newblock IEEE, 2013.

\bibitem{doclo2010acoustic}
Simon Doclo, Sharon Gannot, Marc Moonen, and Ann Spriet,
\newblock ``Acoustic beamforming for hearing aid applications,''
\newblock {\em Handbook on array processing and sensor networks}, pp. 269--302,
  2010.

\bibitem{Sarradj2010}
Ennes Sarradj,
\newblock ``A fast signal subspace approach for the determination of absolute
  levels from phased microphone array measurements,''
\newblock {\em Journal of Sound and Vibration}, vol. 329, no. 9, pp. 1553 --
  1569, 2010.

\end{thebibliography}

\end{document}